# Design of the offline test electronics for the nozzle system of proton therapy

Peng Huang, Zhiguo Yin, Tianjian Bian, Shigang Hou, Yang Wang, Tianjue Zhang, Luyu Ji, Lipeng Wen, Xueer Mu, Rui Xiong

（China Institute of Atomic Energy, 102413, Beijing, China）

## Abstract

A set of nozzle equipment for proton therapy is now being developed at China Institute of Atomic Energy. To facilitate the off-line commissioning of the whole equipment, a set of ionization chamber signal generation system, the test electronics, is designed. The system uses ZYNQ SoC as the main control unit and outputs the beam dose analog signal through DAC8532. The dual SPDT analog switch, DG636, is used to simulate the beam position signals according to Gaussian distribution. The results show that the system can simulate the beam position, dose, and other related analog signals generated by the proton beam when passing through the ionization chamber. Moreover, the accuracy of the simulated beam position is within ± 0.33mm, and the accuracy of the simulated dose signal is within ± 1%. At the same time, it can output analog signals representing environmental parameters. The test electronics meets the design requirements, which can be used to commission the nozzle system as well as the treatment control system without the proton beam.

## 1. Introduction

A proton therapy system based on the 230 MeV superconducting cyclotron is being developed at CIAE[1-4]. The nozzle system, which mainly consists of three ion chambers and two scanning magnets, plays a very important role in the process of Pencil Beam Scanning (PBS). In proton therapy, in order to irradiate tumor lesions according to the preset position and dose, the PBS system is required to accurately control the beam scanning process [5-7]. During spot-scanning irradiation, the ionization chamber (IC) monitors the position and dose of the beam in real time. When the target dose reaches the set value, the beam is turned off and the next spot is irradiated, and so on. When the position/dose error of the beam exceeds the error threshold, the interlock is triggered to turn off the beam. At present, the dose accuracy of the final irradiation of proton therapy devices at home and abroad is generally from ±2% to ±3%, and the position accuracy is generally from ±0.5mm to ±1.5mm [8-11].

Figure 1 shows the structure of the parallel plate ionization chamber [12], in which the strip layer (Axis A strips or Axis B strips) is used to collect the beam position signal, and there are 128 strips in total. The width of a single strip is 1.89mm, and the gap between two adjacent strips is 0.11mm, resulting in an overall spacing of 2mm between strips. The ionization chamber contains two directions X/Y, so there are 256 strips in total. The integral plane (Int plane) is used to collect the beam dose signal. The high voltage layer (HV1 or HV2) is used to provide 2kV positive bias voltage. The ionization chamber is equipped with temperature and pressure sensors, which are used to send environmental information through the Environment monitor readout port to the front-end electronics to calibrate the readout data. The integral plane electrode collects charge over the whole active area of the IC and delivers it to a single readout channel. The strip electrodes partition the measured charge according to where it is formed across the readout area. The partitioning is linear and direct because the field is uniform in the gap. Although the readout strips on the cathodes are separated by very small gaps, all the induced signal and charge eventually arriving at the cathodes is routed by the electric field onto the strips. Figure 2 shows the sequence of electrode foils and the signal partition on strip cathodes.

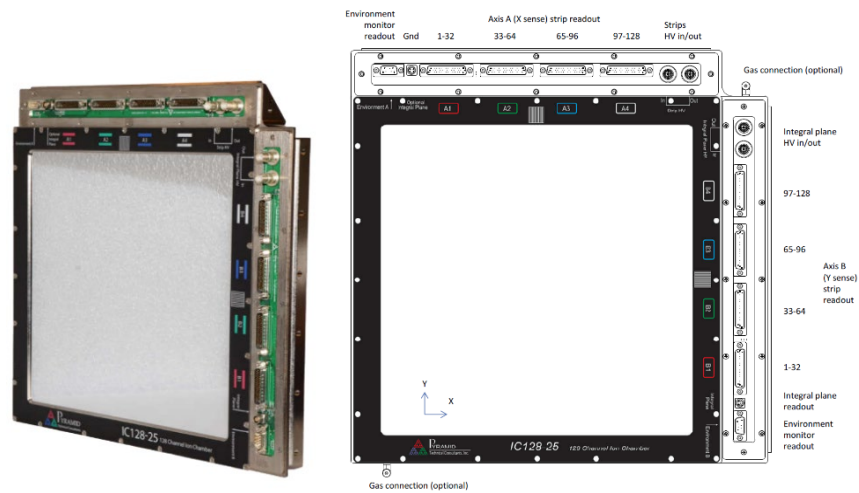

Figure 1 Structure of the parallel plate ionization chamber

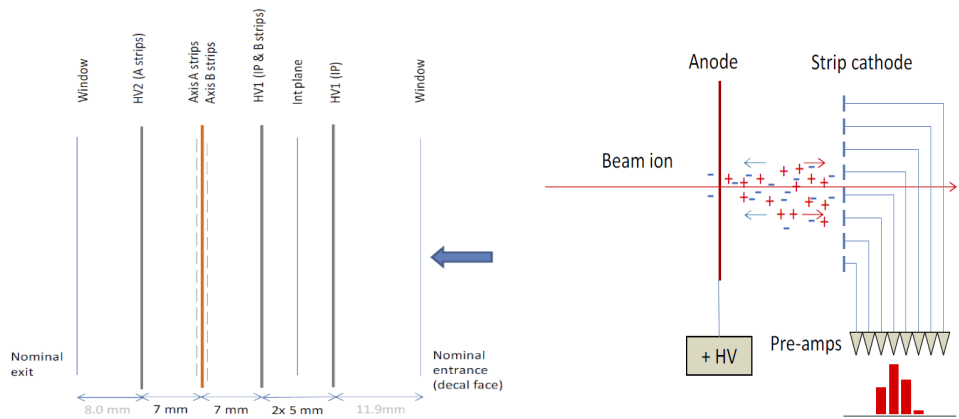

Figure 2 Sequence of electrode foils and signal partition on strip cathodes

So as to conduct testing and upgrades of the on-site nozzle system and its electronics, the beam information is necessary. Also, the beam-related signals are required for running the treatment control system (TCS) [13,14], otherwise, the treatment process cannot be carried out. Therefore, in order to complete the commissioning of the nozzle system and the TCS system, a set of offline test electronics was designed to simulate the signals generated by the IC. In a word, it avoids the dependence on the beam. And after the system is built, it can be used to debug and upgrade the entire treatment facility anytime, anywhere, without the influence of beam quality and accelerator operating state.

## 2. Theoretical design of the test electronics

Since the beam itself exhibits Gaussian distribution [15,16], the signals aroused on the strips after passing through the ionization chamber also exhibit Gaussian distribution, in which the size of the beam spot determines the range (or number) of the strip signals, and the intensity of the beam current determines the amplitude of the strip signals. Figure 3 (a) shows the schematic diagram of the beam passing through the ionization chamber strips (X direction), and Figure 3 (b) shows the signals generated on these strips, which follow the Gaussian distribution as described in Formula 1.

$$f(x) = \frac{1}{\sqrt{2\pi}\sigma} \exp\left(-\frac{(x-\mu)^2}{2\sigma^2}\right) \quad (1)$$

In the formula, μ is the center position of the beam, σ is the standard deviation. The beam position is in the irradiation field (30cm ×40cm), and the beam size (σ) is about 3.2mm~4.4mm for 70~230MeV proton beam.

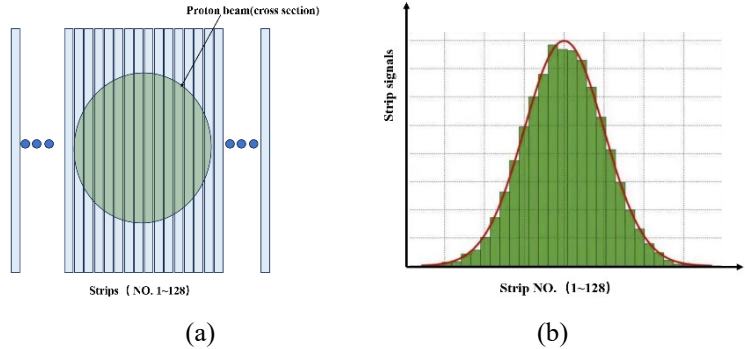

(a) (b)

Figure 3 Beam signals with Gaussian distribution

Therefore, in order to simulate the position signal of the beam on the ionization chamber strips, a conversion algorithm from the actual target position of the beam to the output strip signals is proposed. It takes the beam center position as a reference point and then adopts lookup tables and approximation principles to convert the beam center position coordinates into signal values output on different strips, which is actually equivalent to an approximate inverse operation of a Gaussian distribution. The beam center at any position is ultimately divided into four situations based on the proximity principle. The beam center is in the middle of one strip, in the gap between two strips, in the left quartet of one strip, or in the left quarter of one strip, just as shown in Figure 4 (a)~(d). So, theoretically with this method the maximum deviation can be controlled within ±0.25mm. It is worth noting that in the actual treatment process, the accuracy of the beam position is generally required to be ± 0.5mm~± 1.5mm, so it may meet the needs of offline commissioning.

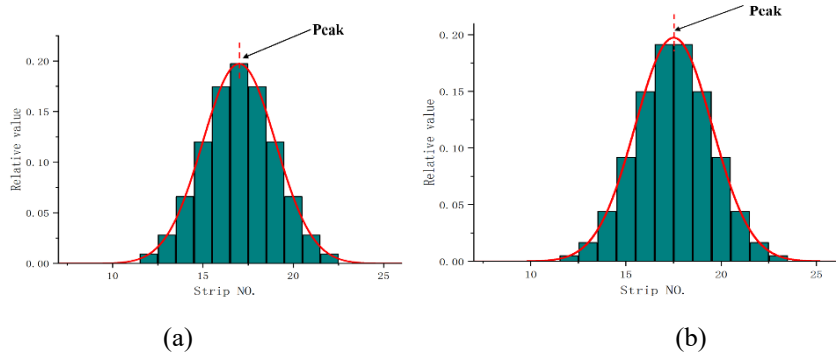

(a) (b)

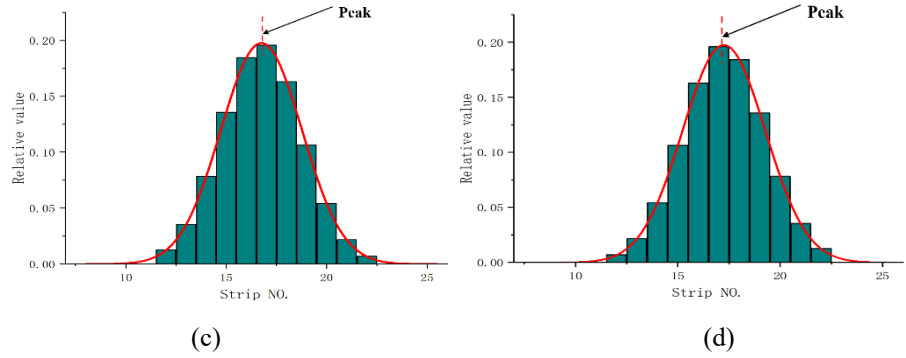

(c)                                        (d)

Figure 4 Schematic diagram of approximation principle (4 conditions)

For the generation of the strip signals, Figure 5 shows the schematic diagram. The design principle of signal output is based on the DAC plus the discharge resistor and the fast switch, which are DAC8532, high precision low temperature drift resistor, and DG636 respectively. The PWM signal from the core control board can adjust the output signal by controlling the duty cycle, as described in the following formula：

$$Q=I*t=(U/R)*t \qquad (2)$$

In the formula, Q is the output charge, I is the output current, t is the on-time of PWM wave, U is the output voltage of DAC, and R is the discharge resistance. It can be seen that different output charge can be realized by controlling the PWM duty cycle (on-time), and the current(I) during this period can also be obtained by the charge and integration time. This is only the control of one output signal, and the control of the overall 128 signals can also be achieved by changing the DAC output voltage U. That is to say, the overall amplitude of the signals is determined by the beam current, and the relative signals on each strip is determined by the coordinates of the beam center. It should be noted that the beam position information (position and σ) fitted from the overall output signals is the most noteworthy, rather than the value and accuracy of the output signal on each channel.

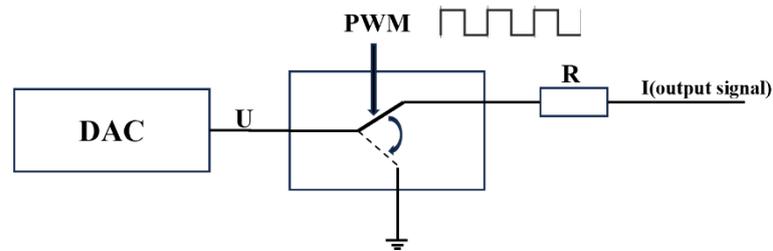

Figure 5   Schematic diagram of the strip signal generation（one channel）

Since the strip signal is generally very small (tens of nanoampere), the injection charge and leakage current of the switch are required to be as small as possible. Also, the turn-on and turn-off time should be as short as possible. Since the integration time of the IC frontend electronics (composed of 128-channel charge integrating amplifiers and ADC) is usually 100μs or more, the PWM period should be set no more than 10μs, at least one tenth of the integration time to ensure the integrity of the integration process and the accuracy and smoothness of the read signals. Finally, the DG636 chip is selected as the switch, whose leakage current is less than 0.5nA, and the turn-on/turn-off time is less than 80ns [17].

The lookup tables for the four situations in Figure 3 are shown in table 1 to 4(σ=4, 100MeV), and the default signal values generated on the other strips is zero. The duty cycle of the strip signal is determined based on the percentage of the peak strip, which is set as 100%. Considering the strip width and beam spot size, according to the 3 Sigma rule, 13 strip signals, a total range of 26 millimeters, are enough to simulate the beam signals.

Table 1 Strip signals distribution (Peak at strip NO.n center)

| Strip NO. | Probability of the beam passing through | (Duty Cycle) Percentage to max value |
|---|---|---|
| n-6 | 0.0024 | 1.2% |
| n-5 | 0.0092 | 4.7% |
| n-4 | 0.028 | 14.2% |
| n-3 | 0.066 | 33.4% |
| n-2 | 0.12 | 60.8% |
| n-1 | 0.17467 | 88.5% |
| **n** | **0.19741** | **100%** |
| n+1 | 0.17467 | 88.5% |
| n+2 | 0.12 | 60.8% |
| n+3 | 0.066 | 33.4% |
| n+4 | 0.028 | 14.2% |
| n+5 | 0.0092 | 4.7% |
| n+6 | 0.0024 | 1.2% |

Table 2 Strip signals distribution (Peak at gap center, strip NO.n-1 and NO.n)

| Strip NO. | Probability of the beam passing through | Percentage to max value (Duty Cycle) |
|---|---|---|
| n-6 | 0.0049 | 2.6% |
| n-5 | 0.0165 | 8.6% |
| n-4 | 0.0441 | 23.0% |
| n-3 | 0.0918 | 47.9% |
| n-2 | 0.1499 | 78.3% |
| **n-1** | **0.1915** | **100%** |
| **n** | **0.1915** | **100%** |
| n+1 | 0.1499 | 78.3% |
| n+2 | 0.0918 | 47.9% |
| n+3 | 0.0441 | 23.0% |
| n+4 | 0.0165 | 8.6% |
| n+5 | 0.0049 | 2.6% |
| n+6 | 0.0011 | 0.6% |

Table 3 Strip signals distribution (Peak at left quarter of strip NO.n)

| Strip NO. | Probability of the beam passing through | Percentage to max value (Duty Cycle) |
|---|---|---|
| n-6 | 0.0034 | 1.7% |
| n-5 | 0.0125 | 6.4% |
| n-4 | 0.0353 | 18.0% |
| n-3 | 0.0782 | 39.9% |
| n-2 | 0.1357 | 69.3% |
| n-1 | 0.1843 | 94.1% |
| **n** | **0.1959** | **100%** |
| n+1 | 0.163 | 83.2% |
| n+2 | 0.1062 | 54.2% |
| n+3 | 0.0542 | 27.7% |
| n+4 | 0.0216 | 11.0% |
| n+5 | 0.0068 | 3.5% |
| n+6 | 0.0017 | 0.9% |

Table 4 Strip signals distribution (d. Peak at right quarter of strip NO.n)

| Strip NO. | Probability of the beam passing through | Percentage to max value (Duty Cycle) |
|---|---|---|
| n-6 | 0.0017 | 0.9% |
| n-5 | 0.0068 | 3.5% |
| n-4 | 0.0216 | 11.0% |
| n-3 | 0.0542 | 27.7% |
| n-2 | 0.1062 | 54.2% |
| n-1 | 0.163 | 83.2% |
| **n** | **0.1959** | **100%** |
| n+1 | 0.1843 | 94.1% |
| n+2 | 0.1357 | 69.3% |
| n+3 | 0.0782 | 39.9% |
| n+4 | 0.0353 | 18.0% |
| n+5 | 0.0125 | 6.4% |
| n+6 | 0.0034 | 1.7% |

As for the amplitude of the dose output signal and the overall tendency of the strip signals, the gain of the ionization chamber and the beam signal itself determine the size of the output signal. As shown in Formula 3, the gain (G) is defined as the output current of the ion chamber ($I_C$) divided by the beam current ($I_B$), which also determines the output value of DAC. Table 5 shows the typical IC gain, the beam size (sigma), and the range in the water of the proton beam, with which the amplitude and range of the IC signals can be derived. Since the air gaps of the strip layer and the Int plane are 7mm and 10mm(2*5mm), respectively, as shown in Figure 1.

$$G = \frac{I_C}{I_B} \qquad (3)$$

Table 5 Typical IC gain, Sigma, and Range in water of the proton beam

| Proton Energy(MeV) | Typical IC gain (mm-1 air gap) | Sigma(mm) | Range in water(cm) |
|---|---|---|---|
| 70 | 30.8 | 4.41 | 4.075 |
| 90 | 25.2 | 4.18 | 6.389 |
| 100 | 23.5 | 4 | 7.707 |
| 120 | 20.5 | 3.79 | 10.65 |
| 150 | 17.5 | 3.53 | 15.76 |
| 180 | 15.4 | 3.38 | 21.63 |
| 200 | 14.4 | 3.31 | 25.93 |
| 230 | 13.5 | 3.28 | 32.91 |

# 3. Design of the hardware circuit

Figure 6 shows the schematic diagram of hardware architecture. The hardware device consists of three parts: the motherboard, the core card, and the 32-channel daughter card. The motherboard integrates with ADCs and DACs, as well as power supply circuits and communication ports. The dose signal circuit, with a structure of DAC-switch-resistor like Figure 4, can generate the dose message representing the Int plane of the ion chamber. Also, the HV sample circuit is used to check the high voltage from the ion chamber and the Env signal circuit is used to offer the environmental signals (temperature and pressure, 0~10v voltage) to the ion chamber for calibration. The core card is developed based on the Xilinx SoC, which also integrates with DDR, FLASH, and other related peripheral circuits. The core card is connected to the motherboard through three high-speed connectors (120Pin*2, 100Pin*1). Four 32-channel daughter cards are used to generate a total of 128 channel signals. Each daughter card contains 32 channel signal output circuits, as shown in Figure 4.

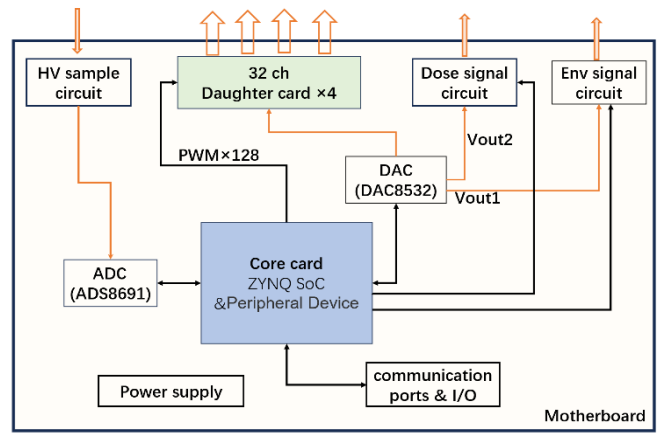

Figure 6 Schematic diagram of hardware architecture

Figure 7 shows the PCB and hardware board of the electronics. For the PCB part, the layout, wiring, and grounding treatment are very important. The XILINX SoC (ZYNQ XC7Z035) [18-21] is selected as the core control unit, which consists of the PS (Processing System) part and the PL (Programmable Logic) part. The PS part is mainly used for communication with the host computer, receiving and sending data, and data preprocessing. The PL part is used for real-time control, such as the PWM signals output and the ADC/DAC driver control. ADAS8691 and DAC8532 are selected as ADC and DAC, respectively, to achieve conversion between the analog and digital quantities. Four 44-pin connectors are used to output the strip signals, and one LEMO connector is used to realize the dose signal output.

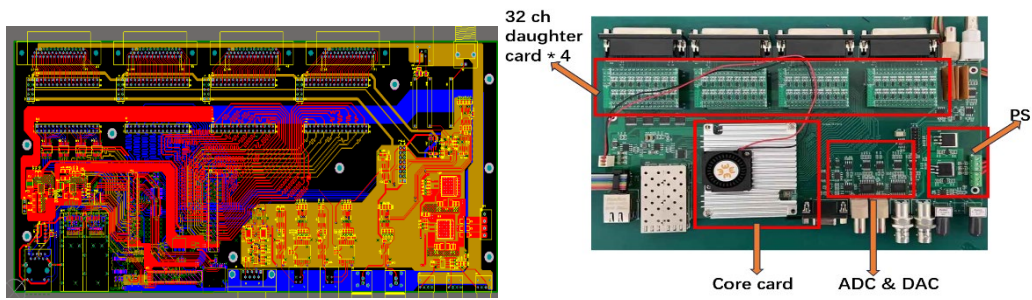

Figure 7 (a) PCB and (b) hardware circuit of the electronics

## 4. Experimental test and results analysis

The testing system consists of three parts: the test electronics, the IC front-end electronics, and the host PC, as shown in Figure 8(a). On the one hand, the host PC sends information related to the irradiation process to the test electronics, and on the other hand, it reads data from the front-end electronics of the ionization chamber. The information mainly includes the target position (x, y), the beam size σ, the target irradiation dose D, the beam current I, beam energy E, etc. Figure 8(b) is the GUI of the test electronics.

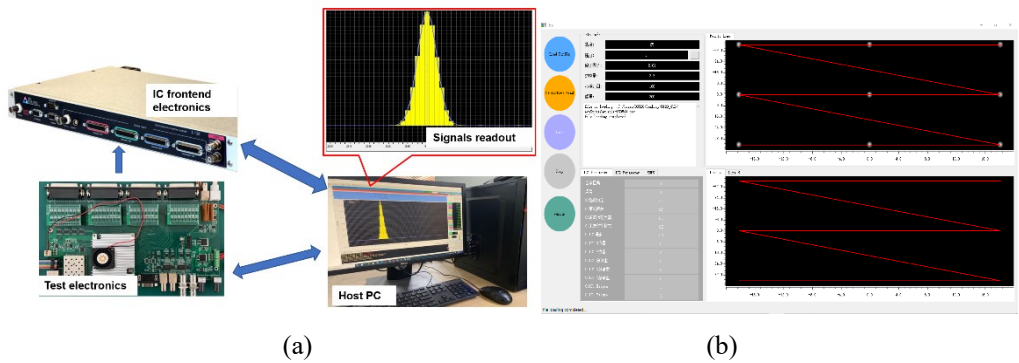

(a) (b)

Figure 8 (a)The testing system and (b) GUI

Firstly, the output of 128 analog signals is tested and the current is measured with Keithley 6517B. Since the gain of the IC strip is about 94~216, then assuming the beam current is 1nA, the total signal generated by the strip layer is 94~216nA. The signal on a single strip generally does not exceed 20% - 30% of the total signal. Therefore, the first two sets of data were measured: 23nA and 39nA. As shown in Figure 9, one of the channels (e.g. ch20) is randomly selected to be measured, with the read current of 23.149 and 38.968nA. At the same time, the measurement results of the 128 channels are shown in Figure 10. On this basis, another two sets of tests were conducted. The first set had a target output value of 80nA, and the 128 signals were measured to be between 79.6nA and 80.5nA. The second target value is 50nA, and the 128 signals were measured to be between 49.4nA and 50.7nA. The differences are mainly related to the accuracy of the selected resistors and the circuit differences of each channel, but the results indicate that the errors are within ± 1nA. Furthermore, just as mentioned above, the experimental results are more concerned with the final position data fitted based on the overall 128 signals, rather than the characteristics of a single simulated strip signal.

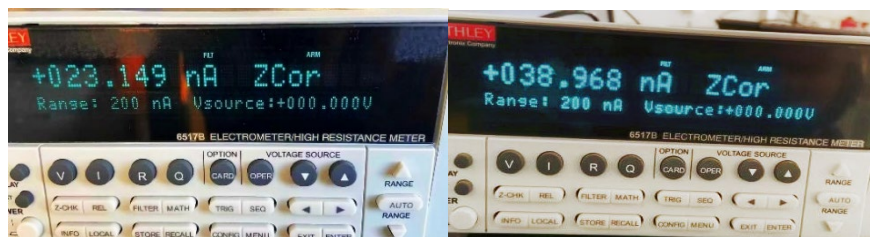

Figure 9 Measurement results of a single output signal

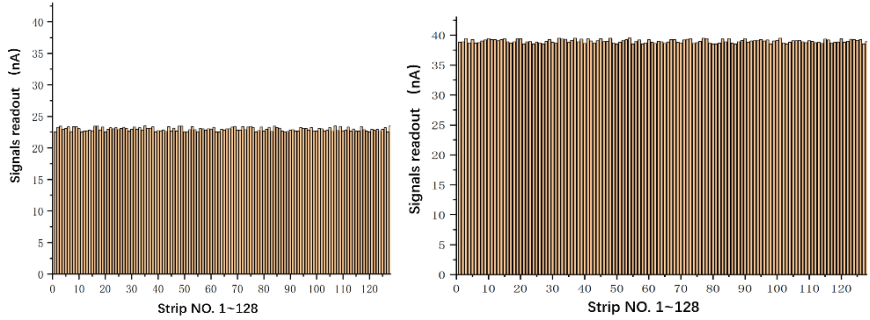

Figure 10 Output of 128 channels

Next, the simulated Gaussian distribution signals are tested. The host PC sends radiation-related data to the offline testing electronics, which then outputs the corresponding analog signals to the front-end electronics of the ionization chamber. Finally, the host PC reads the data back from the front-end electronics. The simulated Gaussian signals and the fitting results are shown in Figure 11, in which (a) ~(d) correspond to four situations in Figure 4. The results, tested under sigma equals 4, show that the output signals follow a Gaussian distribution very well, with the R-square value not less than 0.99.

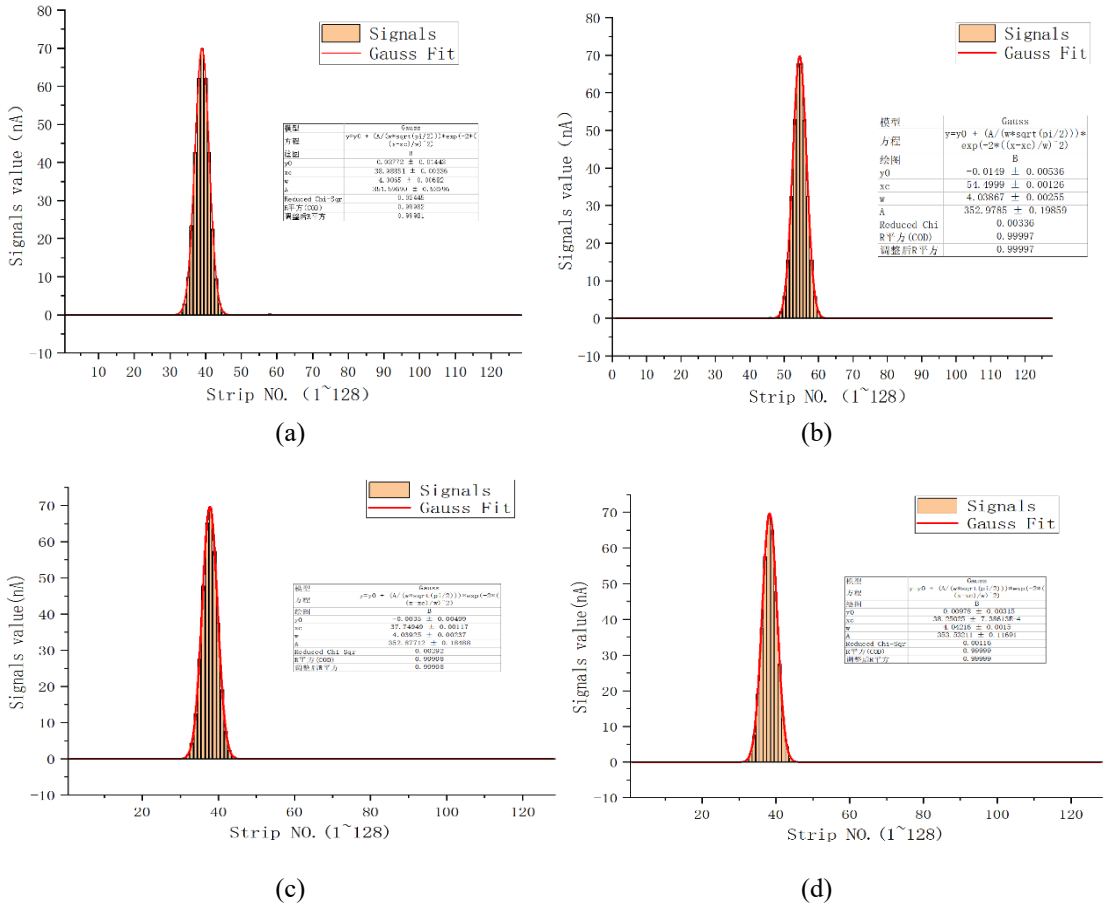

Figure 11 Gaussian signals and fitting results (σ=4)

Furthermore, to convert the fitted strip electrode coordinates into the beam position, the following formulas need to be applied, which are also used to convert the beam position to the strip position by the host PC, then sent to the test electronics.

$$x = (Sx - 64.5) * S \qquad (4)$$

$$y = (Sy - 64.5) * S \qquad (5)$$

Among them, (x, y) is the actual position of the beam. (Sx, Sy) is the strip electrode coordinates fitted based on the simulated signals, and S is the width of the strip spacing, which is 2 mm. Then, according to the same test conditions, the position error experiment was carried out every 4.5mm interval from -100mm to 100mm. The fitted position error is within ±0.067mm, and the sigma error is within ±0.043mm, as shown in Figure 12.

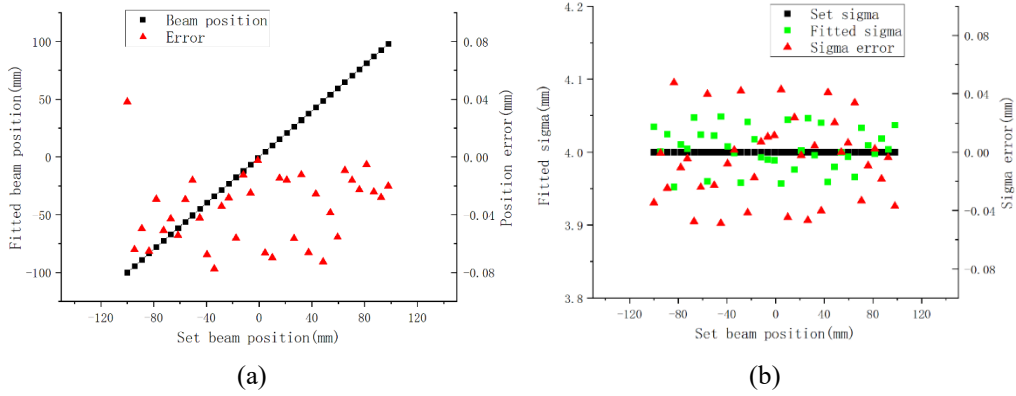

(a) (b)

Figure 12 Position errors and Sigma errors (σ=4)

On this basis, multiple repeated experiments were conducted at different energies (70MeV, 100MeV, 180MeV, 230MeV), and the results showed that the fitted beam center position errors were all within ± 0.08mm, with a maximum sigma error of ± 0.05mm. Combined with the maximum theoretical error of ± 0.25mm, the final position error can be ensured to be within ± 0.33mm, which can fully meet the requirements of general testing occasions.

The output of the beam dose simulation signal and environmental simulation signal is relatively easy and simple. For the dose simulation signals, since the gain of the ionization chamber is generally between 135~612 (70~230MeV), the output analog current is also in the order of 100 nA. So, Six experiments from 100 nA to 600 nA were conducted for this purpose, and the test results show that the accuracy of the simulated dose signals can be easily controlled within ± 1%. Table 6 shows the comprehensive test results with different beam energies.

Table 6 Comprehensive test results

| Item<br>Beam Energy | Beam Position error (fitted+ theoretical error) | Beam Sigma error | R-square | Dose signal output error |
|---|---|---|---|---|
| 70MeV | ±(0.069+0.25) mm | ±0.043 mm | >0.99 | ±1% |
| 100MeV | ±(0.07+0.25) mm | ±0.043 mm | | |
| 180MeV | ±(0.072+0.25) mm | ±0.046 mm | | |
| 230MeV | ±(0.077+0.25) mm | ±0.049 mm | | |
| **70~230MeV** | **±0.33 mm** | **±0.05mm** | **>0.99** | **±1%** |

# 5. Conclusion

In this article, we proposed the test electronics for outputting 128-channel analog signals to simulate the beam position information, and other signals such as the dose signal and the environmental signals are also generated to feed to the IC front-end electronics. With the test electronics, it is possible to test and upgrade the nozzle system and the entire treatment device offline (no proton beam or the accelerator) at any time, without considering the beam quality or the on-site dose after the experiment. It provides a very convenient, fast, and reliable way for the commissioning of the proton therapy facility, also a possibility for the training of operators. The offline test electronics has successfully solved the key problem of generating simulated signals related to the beam, laying a solid foundation for the overall offline test system including the accelerator and the beamline in the future.

## Acknowledgments

This work is supported by the National Natural Science Foundation of China (No. 12205379), and the Leading Innovation Project of China National Nuclear Corporation (No. LC192209000114).